\documentclass[11pt, fceqn]{article}
\usepackage{amsfonts}
\usepackage{flafter}
\usepackage{amssymb,graphics,graphicx,subfigure,caption2,rotating}
\usepackage{amsmath, latexsym}
\usepackage[amsmath,thmmarks]{ntheorem}
\usepackage[top=3.5cm]{geometry}
\begin{document}
\date{}
\title{Nonlinear Improvement of Qubit-qudit Entanglement Witnesses}
\author{Shu-Qian Shen$^1$, Jin-Min Liang$^1$,  Ming Li$^1$, Juan Yu$^1$, Shao-Ming Fei$^{2,3}$\\
{\small{\it $^1$College of Science, China University of Petroleum, 266580 Qingdao, P.R. China}}\\
{\small{\it $^2$Max-Planck-Institute for Mathematics in the Sciences, 04103 Leipzig, Germany}}\\
{\small{\it $^3$School of Mathematical Sciences, Capital Normal University, 100048 Beijing, P.R. China}}}\maketitle
\begin{abstract}
The entanglement witness is an important and experimentally applicable tool for entanglement detection. In this paper, we provide a nonlinear improvement of any entanglement witness for $2\otimes d$ quantum systems. Compared with any existing entanglement witness, the improved separability criterion only needs two more measurements on local observables. Detailed examples are employed to illustrate the efficiency of the nonlinear improvement for general, optimal and non-decomposable entanglement witnesses.

\end{abstract}

\section{Introduction}
As a key physical resource, quantum entanglement can be used to implement many tasks to achieve quantum advantages for quantum communication and computation, see, e.g., \cite{applications}. However, it is extremely hard to determine whether a given quantum state is entangled or not, because this is a NP-hard problem from the viewpoint of computation complexity \cite{Gurvits2003-1}. Up to now, various separable criteria have been suggested, see, e.g., \cite{survey} for comprehensive surveys. However, many of them require density matrix reconstruction via quantum sate tomography, which is challenging in experiments for high dimensional quantum systems.

Entanglement witnesses (EWs) \cite{Terhal2001} provide important feasible approaches for entanglement detection, which do not require the full information of quantum states. An EW is a Hermitian operator, whose mean values for all separable states are nonnegative, but can be negative for at least one entanglement state. It was shown in  \cite{Horodecki1996} that any entanglement state can be detected by at least one EW. Nevertheless, for an unknown entangled state, the construction of the corresponding EW is generally quite difficult. Several constructions of EWs have been proposed for some specific entangled states, see, e.g., \cite{Chruscinski2014,Gerke2018}. EWs can also be used to quantify entanglement \cite{Brandao2005} and design measurement-device-independent ways for entanglement detection \cite{MDIEW}.
The experimental implements of EWs have also been realized in different physical systems \cite{EWphotons}.

The nonlinear improvement of EWs has been addressed based on nonlinear corrections \cite{Guhne2006-1,Guhne2007-1}. In \cite{Moroder2008} the quadratic and nonlinear terms were designed for iteration to find a sequence of stronger nonlinear EWs. Arrazola, Gittsovich and L\"{u}tkenhaus \cite{Arrazola2012} proposed an accessible nonlinear EW, which makes use of the same data as for the evaluation of the original EW. In \cite{Yu2005} a family of EWs based on local orthogonal observables was presented, which has been nonlinearly improved in \cite{Guhne2006-2}. These linear and nonlinear forms have been further optimized in \cite{Zhang2007}.

The separability and entanglement of $2\otimes d$ quantum states have been extensively studied in \cite{2dsystem,Yu2003,Zhao2011}. Specifically, Bell-type inequalities for detecting entanglement were constructed in \cite{Yu2003,Zhao2011}, which provide experimental ways for entanglement detection, as only mean values of local observables are involved.

This paper is devoted to giving an alternative way of general nonlinear improvement of any EW for $2\otimes d$ quantum systems. The inequalities for entanglement detection given in \cite{Yu2003,Zhao2011} can be seen as special examples of our nonlinear improvement of EW. Compared with the original EW, the experimental implementation of the improved criterion only requires two more evaluations of local observables. This is welcome in experimental realization. It should be worth noting that the method used in this paper is completely different from that employed in \cite{Guhne2006-1}-\cite{Arrazola2012}.

The remainder of the paper is organized as follows. After introducing EWs for $2\otimes d$ quantum systems in Section 2, we study the nonlinear improvement of any EW for the $2\otimes d$ quantum system in Section 3, together with detailed examples to illustrate the efficiency of such nonlinear improvement. In Section 4, some concluding remarks are given.

\section{EWs for $2\otimes d$ quantum systems}

Let $A\in \mathbb{C}^{m\times d}$ and $ B\in \mathbb{C}^{d\times d}$. We denote by $A^{\dag}$, $||A||_{\infty},$ $r(B)$, $\gamma(B)$ and $Tr(B)$ the conjugate transpose of $A$, the spectral norm of $A$, the numerical radius of $B$ \cite{Goldberg1982}, the spectral radius of $B$ and the trace of $B$, respectively. The inequality $B>0$ ($B\ge 0$) means that $B$ is Hermitian positive definite (semidefinite).

Any Hermitian operator $W$ acting on $\mathbb{C}^m\otimes \mathbb{C}^d$ is said to be an entanglement witness (EW) if it satisfies the following conditions \cite{Terhal2001}:
\begin{description}
  \item[(i)] $W$ is block positive, i.e., \emph{Tr}$(W\rho_{sep})\ge 0$ for all separable states $\rho_{sep}$ in $ \mathbb{C}^m\otimes \mathbb{C}^d$;
  \item[(ii)] there exists at least an entanglement state $\rho_{ent}$ in $ \mathbb{C}^m\otimes \mathbb{C}^d$ such that \emph{Tr}$(W\rho_{ent})< 0$.
\end{description}

Theoretically, any entangled state can be detected by at least one EW \cite{Horodecki1996}.
The aim of this paper is to improve any given EW for $2\otimes d$ quantum states in terms of a nonlinear improvement.
Any Hermitian operator $W$ acting on $\mathbb{C}^2\otimes \mathbb{C}^d$ can be partitioned into
\begin{equation}
\label{EW} W = \left( {\begin{array}{*{20}{c}}
   {{W_{11}}} & {{W_{12}}}  \\
   {W_{12}^{\dag}} & {{W_{22}}}  \\
\end{array}} \right),
\end{equation}
where $W_{11}$, $W_{22}$ and $W_{12}\in \mathbb{C}^{d\times d}$. We have\\
\\
\textbf{Proposition 1.} \emph{Any Hermitian operator $W$ defined as in \emph{(\ref{EW}}\emph{)} is an EW if and only if the following conditions hold:
\begin{description}
  \item[(a)]$W$ is not positive semidefinite, i.e., $W$ has at least one negative eigenvalue;
  \item[(b)] $W_{11}\ge 0, $ $W_{22}\ge 0$;
  \item[(c)] for any pure state $|\xi\rangle \in \mathbb{C}^d$,
\begin{equation}
\label{EWinequality}
\langle \xi| W_{11} |\xi \rangle \langle \xi| W_{22} |\xi  \rangle\ge  |\langle \xi| W_{12} |\xi\rangle|^2.
\end{equation}
\end{description}}
\smallskip

\noindent\textbf{Proof.} We only need to prove that the item (i) in the definition of EW is equivalent to the items (b)-(c). In fact, the item (i) holds if and only if, for any pure states $|\phi\rangle \in \mathbb{C}^2$ and $ |\xi\rangle \in \mathbb{C}^d$,
\[
\langle \phi,\xi|W|\phi,\xi\rangle = \langle \phi|\left(\langle \xi|W|\xi\rangle\right)|\phi\rangle = \langle\phi|\left( {\begin{array}{*{20}{c}}
   {{\langle \xi |W_{11}|\xi \rangle}} & {{\langle \xi |W_{12}|\xi \rangle}}  \\
   {\langle \xi |W_{12}^{\dag}|\xi \rangle} & {{\langle \xi |W_{22}|\xi \rangle}}  \\
\end{array}} \right)|\phi\rangle\ge 0,
\]
i.e., for any pure state $|\xi\rangle\in\mathbb{C}^d$, $\langle\xi|W|\xi \rangle\ge 0$. By \cite{Horn1985}, the inequality $\langle\xi|W|\xi \rangle\ge 0$, $\forall |\xi\rangle\in\mathbb{C}^d$, is further equivalent to the fact that the inequalities $\langle\xi|W_{11}|\xi \rangle\ge 0$, $\langle\xi|W_{22}|\xi \rangle\ge 0$ and (\ref{EWinequality}) hold for any pure state $ |\xi\rangle\in\mathbb{C}^d$.
$\hfill\Box$

Up to now there are few constructions of EWs for $2\otimes d$ quantum systems when $d>2$. In particular, if $W_{22}=\alpha W_{11}>0$ for some $\alpha>0$, then the item (a) in Proposition 1 holds if and only if the Schur complement of $W_{11}$,
\[
W_{22}-W_{12}^{\dag}W_{11}^{-1}W_{12}= W_{11}^{\frac{1}{2}}(\alpha I_d- W_{11}^{-\frac{1}{2}}W_{12}^{\dag}W_{11}^{-1}W_{12} W_{11}^{-\frac{1}{2}})W_{11}^{\frac{1}{2}},
\]
is not positive semidefinite \cite{Horn1985}, i.e.,
\[
\gamma\left(W_{11}^{-\frac{1}{2}}W_{12}^{\dag}W_{11}^{-1}W_{12} W_{11}^{-\frac{1}{2}}\right)=||\tilde{W}_{12}||_{\infty}^2>\alpha,
\]
where $\tilde{W}_{12}=W_{11}^{-\frac{1}{2}}W_{12}W_{11}^{-\frac{1}{2}}$, and $I_d$ is the $d\times d$ identity matrix.
While the item (c) in Proposition 1 holds if and only if for any pure state $|\xi\rangle\in \mathbb{C}^d$,
\[
\frac{|\langle \xi|W_{12}|\xi\rangle|^2}{\langle \xi| W_{11}|\xi\rangle^2}=\frac{|\langle \eta|\tilde{W}_{12}|\eta\rangle|^2}{\langle \eta|\eta\rangle^2}\le \alpha,
\]
which is equivalent to $r(\tilde{W}_{12})\le \sqrt{\alpha}$, where $|\eta\rangle=W_{11}^{\frac{1}{2}}|\xi\rangle$ may be not normalized.
Therefore, if $W_{22}=\alpha W_{11}>0$ for some $\alpha>0$, $W$ is an EW if and only if
\begin{equation}\label{t32}
r(\tilde{W}_{12})\le \sqrt{\alpha}<||\tilde{W}_{12}||_{\infty}.
\end{equation}

\section{Nonlinear improvement of any $2\otimes d$ EW}

Let the Hermitian operator $W$ be defined as in (\ref{EW}). We define
\begin{equation}
\label{OEW1} W_1 = \left( {\begin{array}{*{20}{c}}
   {{W_{11}}} & {0}  \\
   {0} & {{0}}  \\
\end{array}} \right),~~W_2= \left( {\begin{array}{*{20}{c}}
   {{0}} & {0}  \\
   {0} & {{W_{22}}}  \\
\end{array}} \right)~ \text{ and }~W_3=\left( {\begin{array}{*{20}{c}}
   {{0}} & {W_{12}}  \\
   {W_{12}^{\dag}} & {{0}}  \\
\end{array}} \right).
\end{equation}
\smallskip
\\
Clearly, $W_1$ and $W_2$ can be locally decomposed into
 \begin{equation}
 \label{Wdecom}W_1=|0\rangle\langle 0|\otimes W_{11},~~~ W_2=|1\rangle\langle 1|\otimes W_{22}.
 \end{equation}
 The following proposition establishes an equivalent condition of $W$ being block positive.\\
\\
\textbf{Proposition 2.} \emph{Let $W_{11}\ge 0$ and $ W_{22}\ge 0$. Then $W$ is block positive if and only if, for any separable state $\rho_{sep}$ in $ \mathbb{C}^2\otimes \mathbb{C}^d$,
\begin{equation}
\label{pro31}\langle W_1\rangle_{\rho_{sep}}\langle W_2\rangle_{\rho_{sep}}\ge \frac{1}{4}|\langle W_3\rangle_{\rho_{sep}}|^2.
\end{equation}
  }
\noindent\textbf{Proof.} From Proposition 1, $W$ is block positive if and only if the inequality (\ref{EWinequality}) holds for any $|\xi\rangle\in \mathbb{C}^d$.
We now prove (\ref{EWinequality}) is equivalent to (\ref{pro31}).

(\ref{EWinequality})$\Longrightarrow$(\ref{pro31}): We first consider the case when $\rho_{sep}$ is a separable pure state: $\rho_{sep}=|\eta\rangle\langle \eta|$ with
\begin{equation}
\label{pro32}
|\eta\rangle=|\phi\rangle\otimes |\xi\rangle,~~~|\phi\rangle=\left( {\begin{array}{*{20}{c}}
   {a_0}  \\
   {a_1}  \\
\end{array}} \right).
\end{equation}
One has
\begin{align}
\label{Pro33-1}
\langle W_{1}\rangle_{|\eta\rangle\langle \eta|}\langle W_{2}\rangle_{|\eta\rangle\langle \eta|}&=|a_0|^2|a_1|^2\langle \xi|W_{11}|\xi\rangle \langle \xi|W_{22}|\xi\rangle\\
\label{Pro33-2}&\ge |\text{Re}(a_0^*a_1\langle \xi|W_{12}|\xi\rangle)|^2
=\frac{1}{4}|\langle W_3\rangle_{|\eta\rangle \langle \eta|}|^2,
\end{align}
where $*$ denotes the conjugate of a complex number, and we have used (\ref{EWinequality}) for the inequality. Thus, the inequality (\ref{pro31}) holds for any separable pure state.

Any separable mixed state $\rho_{sep}$ can be decomposed into
$
\rho_{sep}=\sum\nolimits_{i}p_i|\tau_i\rangle\langle \tau_i|,
$
where $|\tau_i\rangle=|\phi_i\rangle\otimes |\xi_i\rangle, p_i>0,\sum\nolimits_{i}p_i=1$. Set
\[
A=\frac{1}{2}(W_1+W_2),~~~B=\frac{1}{2}(W_1-W_2).
\]
From (\ref{Pro33-1}) and (\ref{Pro33-2}), we have, for any $|\tau_i\rangle$,
\begin{equation*}
\langle W_1\rangle_{|\tau_i\rangle\langle\tau_i|}\langle W_2\rangle_{|\tau_i\rangle\langle\tau_i|}=\langle A\rangle_{|\tau_i\rangle\langle\tau_i|}^2-\langle B\rangle_{|\tau_i\rangle\langle\tau_i|}^2\ge \frac{1}{4}|\langle W_3\rangle_{|\tau_i\rangle\langle \tau_i|}|^2,
\end{equation*}
which implies
\begin{equation*}
\langle A\rangle_{|\tau_i\rangle\langle\tau_i|}^2\ge \langle B\rangle_{|\tau_i\rangle\langle\tau_i|}^2+ \frac{1}{4}|\langle W_3\rangle_{|\tau_i\rangle\langle \tau_i|}|^2.
\end{equation*}
From Lemma 1 in \cite{Zhao2011}, we obtain
 \begin{equation*}
 \langle A\rangle_{\rho_{sep}}^2\ge \langle B\rangle_{\rho_{sep}}^2+ \frac{1}{4}|\langle W_3\rangle_{\rho_{sep}}|^2,
 \end{equation*}
which is equivalent to (\ref{pro31}).

(\ref{pro31})$\Longrightarrow$(\ref{EWinequality}): For any pure separable state $|\eta\rangle$ defined as in (\ref{pro32}), we obtain, from (\ref{pro31}),
\begin{align*}
\langle W_1\rangle_{|\eta\rangle\langle\eta|}\langle W_2\rangle_{|\eta\rangle\langle\eta|}&=|a_0|^2|a_1|^2\langle\xi|W_{11}|\xi\rangle\langle\xi|W_{22}|\xi\rangle\\
&\ge \frac{1}{4}|\langle W_3\rangle_{|\eta\rangle\langle\eta|}|^2=|\text{Re}(a_0^*a_1\langle\xi|W_{12}|\xi\rangle)|^2.
\end{align*}
Assume that $
\langle\xi|W_{12}|\xi\rangle=x+\textbf{\text{i}}y,x,y\in\mathbb{R},$
where $\textbf{\text{i}}=\sqrt{-1}$. Then the above inequality can be further written as
$
\langle\xi|W_{12}|\xi\rangle=\sqrt{x^2+y^2}(\cos\theta+\textbf{\text{i}}\sin\theta),
$
where $\theta$ is the argument of  $\langle\xi|W_{12}|\xi\rangle$ satisfying $-\pi< \theta\le \pi$.
If we choose
\[
a_0=\frac{1}{\sqrt{2}},a_1=\frac{1}{\sqrt{2}}(\cos\theta-\textbf{\text{i}}\sin\theta),
\]
then the inequality (\ref{EWinequality}) holds. $\hfill\Box$

Based on Proposition 2, we now give the main result.\\
\\
\textbf{Corollary 1.} \emph{Let $W$ defined as in \emph{(}\emph{\ref{EW}}\emph{)} be an EW.
Then any separable state $\rho_{sep}$ in $\mathbb{C}^2\otimes \mathbb{C}^d$ satisfies
\begin{equation}\label{thm}
\langle W_1\rangle_{\rho_{sep}}\langle W_2\rangle_{\rho_{sep}}\ge \frac{1}{4}|\langle W_3\rangle_{\rho_{sep}}|^2.
\end{equation}}

The mean value of $W_3=W-W_1-W_2$ can be obtained from the mean values of $W$, $W_1$ and $W_2$. Therefore, compared with the witness $W$, to use (\ref{thm}) to detect entanglement, one needs only two more measurements on the local observables $W_1$ and $W_2$ given in (\ref{Wdecom}).
Thus, the inequality (\ref{thm}) is very experimentally plausible for entanglement detection.

We now show that the inequality (\ref{thm}) is
better than the witness $W$ itself in entanglement detection.
\\
\\
\textbf{Proposition 3.} \emph{Let $W$ defined as in \emph{(}\emph{\ref{EW}}\emph{)} be an EW. Any state $\rho$ with $\langle W\rangle_{\rho}<0$ can be detected by }(\ref{thm}).
\\
\\
\textbf{Proof.} For any state $\rho$ in $\mathbb{C}^2\otimes \mathbb{C}^d$, if
\begin{equation}\label{pr331p}
\langle W_1\rangle_{\rho}\langle W_2\rangle_{\rho}\ge\frac{1}{4}|\langle W_3\rangle_{\rho}|^2,
\end{equation}
then one gets
\begin{align*}
\langle W\rangle_{\rho}&=\langle W_1\rangle_{\rho}+\langle W_2\rangle_{\rho}+\langle W_3\rangle_{\rho}\ge 2\sqrt{\langle W_1\rangle_{\rho}\langle W_2\rangle_{\rho}}+\langle W_3\rangle_{\rho}\\
&\ge |\langle W_3\rangle_{\rho}|+\langle W_3\rangle_{\rho}\ge 0.
\end{align*}
Thus, any entangled state $\rho$ with $\langle W\rangle_{\rho}<0$ must violate the inequality (\ref{pr331p}), and then can be detected by (\ref{thm}).
$\hfill\Box$

Hence, any $2\otimes d$ EW can be improved by our nonlinear inequality (\ref{thm}) in Corollary 1.
The nonlinear improvements of any EW have also been studied in \cite{Guhne2006-1}-\cite{Arrazola2012}, which are completely different from that given by (\ref{thm}). In fact, their improvements depend on  nonlinear corrections by adding quadratic and nonlinear terms to the linear EW, while the improvement by (\ref{thm}) is based on a nonlinear inequality with the decomposition of the linear EW: $W=W_1+W_2+W_3$.
Take \cite{Guhne2006-1}, for instance. Let an EW $W$ has the form $W=|\phi\rangle\langle \phi|^{T_2}$, where $T_2$ denotes the partial transpose with respect to the second subsystem. Then Observation 1(a) in \cite{Guhne2006-1} shows that
\begin{equation}\label{Guhne}
\mathcal{G}(\rho)=\langle W\rangle_{\rho}-\frac{1}{s(\xi)}\langle V^{T_2}\rangle_{\rho}\langle \left(V^{T_2}\right)^\dag\rangle_{\rho}
\end{equation}
is a nonlinear improvement of $W$, where $V=|\phi\rangle\langle \xi|$ for any $|\xi\rangle$, and $s(\xi)$
is the square of the largest Schmidt coefficient of $|\xi\rangle$. Consider the $2\otimes 2$ state
\[
|\eta\rangle=\frac{1}{\sqrt{2}}(|01\rangle-|10\rangle).
\]
If we choose
\[
|\phi\rangle=\frac{1}{\sqrt{2}}(|00\rangle-|11\rangle),
\]
then $W=|\phi\rangle\langle \phi|^{T_2}$ is an EW. For any $|\xi\rangle=(x_1,x_2,x_3,x_4)^\dag$, simple computation yields $\langle W\rangle_{|\eta\rangle\langle \eta|}=\frac{1}{2}>0$, and
\[
\mathcal{G}(|\eta\rangle \langle \eta|)=\frac{1}{2}-\frac{|x_1-x_4|^2}{4f(x_1,x_2,x_3,x_4)}> 0,
\]
where
\[
f(x_1,x_2,x_3,x_4)=1+\sqrt{\left(|x_1|^2-|x_2|^2+|x_3|^2-|x_4|^2\right)^2+4|\bar{x}_1x_2+\bar{x}_3x_4|^2}.
\]
Thus, $W$ and (\ref{Guhne}) cannot detect entanglement in $|\eta\rangle$, but the inequality (\ref{thm}) can detect it by
\[
\langle W_1\rangle_{|\eta\rangle\langle \eta|}\langle W_2\rangle_{|\eta\rangle\langle \eta|}-\frac{1}{4}|\langle W_3\rangle_{|\eta\rangle\langle \eta|}|^2=-\frac{1}{16}<0.
\]

The following proposition shows that the condition $W$ being an EW cannot be weakened.
\\
\\
\textbf{Proposition 4.} \emph{Let $W_{11}\ge 0$ and $W_{22}\ge 0$. Then
   $W\ge 0$ if and only if, for any state $\rho$ in $\mathbb{C}^2\otimes \mathbb{C}^d$,
\begin{equation}
\label{pr331}
\langle W_1\rangle_{\rho}\langle W_2\rangle_{\rho}\ge\frac{1}{4}|\langle W_3\rangle_{\rho}|^2.
\end{equation}}
\textbf{Proof.}
(\ref{pr331}) $\Longrightarrow$ $W\ge 0$ : From the proof of Proposition 3, for any state $\rho\in\mathbb{C}^2\otimes \mathbb{C}^d$, the inequality (\ref{pr331}) leads to $\langle W\rangle_{\rho}\ge 0$, and then $W\ge 0$.

$ W\ge 0$ $\Longrightarrow$ (\ref{pr331}): We consider the following two cases.

\emph{Case \emph{1}.} $W_{11}> 0$. In this case, any pure state $|\phi\rangle\in \mathbb{C}^2\otimes \mathbb{C}^d$ can be partitioned into
$
|\phi\rangle=\left( {\begin{array}{*{20}{c}}
   {|\phi_1\rangle}  \\
   {|\phi_2\rangle}  \\
\end{array}} \right),
$
where $|\phi_i\rangle\in \mathbb{C}^d$, $i=1,2$, may be not normalized. We get
\begin{align*}
|\langle W_3\rangle_{|\phi\rangle\langle \phi|}|^2&=4|\text{Re}(\langle \phi_1|W_{12}|\phi_2\rangle)|^2\le 4|\langle \phi_1|W_{12}|\phi_2\rangle|^2\nonumber \\
\label{NEWin}&\le 4\langle \phi_1|W_{11}|\phi_1\rangle\langle \phi_2|W_{12}^{\dag}W_{11}^{-1}W_{12}|\phi_2\rangle\\
&\le 4\langle \phi_1|W_{11}|\phi_1\rangle\langle \phi_2|W_{22}|\phi_2\rangle=4\langle W_1\rangle_{|\phi\rangle\langle\phi|}\langle W_2\rangle_{|\phi\rangle\langle\phi|},
\end{align*}
where we have used the Cauchy-Schwarz inequality in the second inequality.  The third inequality  is due to the fact that, for the case $W_{11}>0$, $W$ is positive semidefinite if and only if $W_{22}-W_{12}^{\dag}W_{11}^{-1}W_{12}$ is positive semidefinite \cite{Horn1985}. Therefore, the inequality (\ref{pr331}) holds for any pure state.

For any mixed state $\rho$ with spectral decomposition, $\rho=\sum\nolimits_{i}p_i|\tau_i\rangle\langle \tau_i|$, $p_i\ge 0$, $\sum\nolimits_{i}p_i=1$, by an analogous argument as in the proof of ``(\ref{EWinequality})$\Longrightarrow$(\ref{pro31})" from Proposition 2, we get (\ref{pr331}) for any mixed state.

\emph{Case \emph{2}.} $W_{11}\ge 0$. Denote $W(\epsilon)=W+\epsilon I_{2d}$ for any positive number $\epsilon$. From $W(\epsilon)>0$ and \emph{Case} 1, we have
\begin{equation*}
\langle W_1(\epsilon)\rangle_{\rho}\langle W_2(\epsilon)\rangle_{\rho}\ge\frac{1}{4}|\langle W_3(\epsilon)\rangle_{\rho}|^2,
\end{equation*}
where $W_1(\epsilon)$, $W_2(\epsilon)$ and $W_3(\epsilon)$ for $W(\epsilon)$ are analogously defined as in (\ref{OEW1}). Thus, as $\epsilon\rightarrow 0^+$, we can get (\ref{pr331}).
$\hfill \Box$

Thus, the inequality (\ref{thm}) in Corollary 1 cannot detect any entanglement when $W$ is not an EW.
 The following simple example shows that the optimal EW \cite{Lewenstein2000} can be improved by our criterion (\ref{thm}).
\\
\\
\textbf{Example 1.} Consider the mixture of a Bell state with white noise
\[
\rho_p=p|\phi\rangle\langle\phi|+\frac{1-p}{4}I_4,
\]
where
\[
|\phi\rangle=\frac{1}{\sqrt{2}}(|01\rangle+|10\rangle),~~0\le p\le 1.
\]
The flip operator $F=\sum\nolimits_{i,j=0}^{1}|i\rangle\langle j|\otimes |j\rangle\langle i|$
is an optimal EW \cite{Chruscinski2014}. Numerical computation shows that $F$ cannot detect any entanglement of $\rho_p$. But the nonlinear improvement of $F$ from (\ref{thm}) can detect the entanglement of $\rho_p$ for $0.3334\le p\le 1$.

We now construct a general EW based on (\ref{t32}). Consider the following Hermitian operator acting on $\mathbb{C}^2\otimes \mathbb{C}^4$:
\begin{equation*}
\label{example1} W_s^{\alpha}= \left( {\begin{array}{*{20}{c}}
   {{I_4}} & {{W_{12}}}  \\
   {W_{12}^{\dag}} & {{\alpha I_4}}  \\
\end{array}} \right),
\end{equation*}
where
\[
W_{12}=\left( {\begin{array}{*{20}{c}}
   {\rm{0}} & {\frac{{\rm{3}}}{{\rm{2}}}} & {\rm{0}} & {\rm{0}}  \\
   {\rm{0}} & {\rm{0}} & {\rm{0}} & {\rm{0}}  \\
   {\rm{0}} & {\rm{0}} & {\rm{0}} & {\frac{{\rm{3}}}{{\rm{2}}}}  \\
   {\rm{0}} & {\rm{0}} & {\rm{0}} & {\rm{0}}  \\
\end{array}} \right).
\]
If we take $\alpha=r(W_{12})^2=0.75^2$, then from (\ref{t32}), $W_s^{\alpha}$ is an EW by $||W_{12}||_\infty=1.5$. The following example verifies the corresponding improvement of $W_s^{\alpha}$ from (\ref{thm}).
\\
\\
\textbf{Example 2.} Consider the following $2\otimes 4$ state
\begin{equation*}
\rho_p=p|\xi\rangle\langle\xi|+(1-p)\rho,
\end{equation*}
where $0\le p \le 1$, $|\xi\rangle=\frac{1}{\sqrt{2}}(|00\rangle+|11\rangle)$, and
$\rho$ is the positive partial transpose (PPT) entangled state constructed in \cite{Horodecki1997}:
\begin{equation}
\label{ppts}\rho=\frac{1}{{7b + 1}}\left( {\begin{array}{*{20}{c}}
   b & 0 & 0 & 0 & 0 & b & 0 & 0  \\
   0 & b & 0 & 0 & 0 & 0 & b & 0  \\
   0 & 0 & b & 0 & 0 & 0 & 0 & b  \\
   0 & 0 & 0 & b & 0 & 0 & 0 & 0  \\
   0 & 0 & 0 & 0 & {\frac{1}{2}(1{\rm{ + \emph{b})}}} & 0 & 0 & {\frac{1}{2}\sqrt {1 - {b^2}} }  \\
   b & 0 & 0 & 0 & 0 & b & 0 & 0  \\
   0 & b & 0 & 0 & 0 & 0 & b & 0  \\
   0 & 0 & b & 0 & {\frac{1}{2}\sqrt {1 - {b^2}} } & 0 & 0 & {\frac{1}{2}(1{\rm{+\emph{b})}}}  \\
\end{array}} \right),
\end{equation}
with $0<b<1$.

Table \ref{tableZhao} displays the entanglement detection of $\rho_p$ by $W_s^{\alpha}$ and its nonlinear improvement from (\ref{thm}) with $\alpha=0.75^2$. It can be found that the improved criteria from our criterion (\ref{thm}) is more efficient than $W_s^{\alpha}$.

\begin{table}[htbp]
\centering \begin{tabular} {c|c|c|c|c}\hline
{}&$b=0.2$ & $b=0.4$ &$b=0.6$ &$b=0.8$
 \\ \hline
{$W_s^{\alpha}$}&$0.2248\le p\le 1$ & $0.1381\le p\le 1$ &$0.0912\le p\le 1$ &$0.0619\le p\le 1$
 \\ \hline
  {(\ref{thm}) from $W_s^{\alpha}$}&$0.2246\le p\le 1$ & $0.1282\le p\le 1$ &$0.0691\le p\le 1$ &$  0.0290\le p\le 1$
  \\ \hline 
\end{tabular} \caption{\emph{Entanglement detection of $\rho_p$ for different values of $b$.}}
\label{tableZhao}
\end{table}
The following example shows that the non-decomposable EWs \cite{Lewenstein2000} can also be strictly improved by our criterion (\ref{thm}).
\\
\\
\textbf{Example 3.} The PPT entanglement state $\rho$ defined in (\ref{ppts}) can be transformed into the other PPT entanglement state:
\[
\tilde\rho_b=I_2\otimes V\rho I_2\otimes V^\dag,
\]
 where
\[
V=\frac{1}{\sqrt{2}}[(I_2+\textbf{\text{i}}\sigma_1)_{03}\oplus (I_2+\textbf{\text{i}}\sigma_1)_{12}],
\]
where the sub-indices $03$ and $12$ denote the subspaces of the four dimensional space.

 Based on $\tilde\rho_b$, Lewenstein et al. \cite{Lewenstein2000} constructed the non-decomposable EWs in detecting the entanglement of $\tilde \rho_b$. Let $P_b$ be the projector onto the kernel of $\tilde\rho_b$. Then $W^L_b=P_b+P_b^{T_2}-\epsilon_b I_8$ is a non-decomposable EW, where $T_2$ denotes the partial transpose with respect to the second system, and
\[
\epsilon_b=\inf\limits_{|e,f\rangle}\langle e,f|(P_b+P_b^{T_2})|e,f\rangle.
\]
We now use $W^L_b$ and its nonlinear improved criterion (\ref{thm}) to detect the entanglement in the sate
$\tilde \rho_b^p=p\tilde\rho_b+\frac{1-p}{8}I_8$. Table \ref{table2} shows the entanglement conditions of $\tilde \rho_b^p$ from  $W^L_b$ and its nonlinear improved criterion based on (\ref{thm}) for different values of $b$. It is easy to see that the EW $W_b^L$ is always weaker than the corresponding nonlinear improved criterion from (\ref{thm}).

\begin{table}[htbp]
\centering \begin{tabular} {c|c|c}\hline
{}&$W_b^{L}$ & (\ref{thm}) from $W_b^{L}$
 \\ \hline
 {$b=0.2$}&$\small{0.98073582801\le p\le 1}$ & $ 0.98072097191\le p\le 1$
 \\ \hline
    {$b=0.4$}&$ 0.98681160139 \le p\le 1$ & $ 0.98681096542\le p\le 1$
 \\ \hline
        {$b=0.6$}&$0.99264908563\le p\le 1$ & $0.99264906811\le p\le 1$
 \\ \hline    {$b=0.8$}&$0.99698286647\le p\le 1$ & $0.99698286637\le p\le 1$
 \\ \hline
\end{tabular} \caption{\emph{Entanglement conditions of $\tilde \rho_b^p$ from $W_b^{L}$ and its nonlinear improvement based on} (\ref{thm}).}
\label{table2}
\end{table}

We have shown by detailed examples that the nonlinear improved criterion (\ref{thm}) of any EW $W$ is always better than $W$ itself in quantum entanglement detection. As a last remark, let us consider the following operator acting on $\mathbb{C}^2\otimes \mathbb{C}^d$,
\begin{equation*}\label{ZhaoEW}
W_z^{U,V}=U\otimes V \bar W_z U^{\dag}\otimes V^\dag,
\end{equation*}
where
\[
\bar W_z =\frac{1}{2} \left( {\begin{array}{*{20}{c}}
   {{|0\rangle \langle 0|}} & |1\rangle \langle 0|  \\
   |0\rangle \langle 1| & |1\rangle\langle 1|  \\
\end{array}} \right),
\]
$U$ and $V$ are $2\times 2$ and $d\times d$ unitary matrices, respectively.
Since for any pure state $|\xi\rangle=(x_1,x_2,\cdots,x_d)^\dag$,
$
\langle \xi|0\rangle \langle 0|\xi\rangle \langle \xi|1\rangle\langle 1|\xi\rangle -\langle \xi|1\rangle \langle 0|\xi\rangle \langle \xi|0\rangle\langle 1|\xi\rangle=0,
$
$ W^{U,V}_z$ is block positive. Meanwhile, one eigenvalue of $\bar W_z$ is $-\frac{1}{2}$. Hence $W_z^{U,V}$is an EW.

Concerning the nonlinear improvement of $W_z^{U,V}$, let $A_i=U\sigma_iU^{\dag}$ be observables given by Pauli matrices $\sigma_i$, $i=1,2,3$. Set $B_j=V\lambda_jV^{\dag}$, $j=1,\cdots,d+1$, where
\begin{align*}
&\lambda_1=|0\rangle\langle 0|-|1\rangle\langle 1|,~~~\lambda_2=|0\rangle\langle 0|-|2\rangle\langle 2|,\cdots,\\
&\lambda_{d-1}=|0\rangle\langle 0|-|d-1\rangle\langle d-1|,~~~\lambda_d=|0\rangle\langle 1|+|1\rangle\langle 0|,\\
&\lambda_{d+1}=\text{\textbf{i}}|0\rangle\langle 1|-\text{\textbf{i}}|1\rangle\langle 0|.
\end{align*}
Denote
\begin{align*}
&R_1=2I_2\otimes I_d+(2-d)I_2\otimes B_1+2I_2\otimes B_2+\cdots+2I_2\otimes B_{d-1}+dA_3\otimes B_1,\\
&R_2=dI_2\otimes B_1+2A_3\otimes I_d+(2-d)A_3\otimes B_1+2A_3\otimes B_2+\cdots+2A_3\otimes B_{d-1},\\
&R_3=2d (A_1\otimes B_d+A_2\otimes B_{d+1}).
\end{align*}
Then the EW $W_z^{U,V}$ can be exactly written as
\[
W_z^{U,V}=\frac{1}{Tr(2R_1+R_3)}((R_1+R_2)+(R_1-R_2)+R_3)=\frac{1}{Tr(2R_1+R_3)}(2R_1+R_3).
\]
From  (\ref{thm}) in Corollary 1, we have, for any separable state $\rho_{sep}$ in $\mathbb{C}^2\otimes \mathbb{C}^d$,
\[
\langle R_1+R_2\rangle_{\rho_{sep}}\langle R_1-R_2\rangle_{\rho_{sep}}\ge \frac{1}{4}\langle R_3\rangle^2_{\rho_{sep}},
\]
which can be further written as
\begin{equation}
\label{Zhaoin}
\langle R_1\rangle_{\rho_{sep}}\ge \left(\langle R_2\rangle_{\rho_{sep}}^2+\frac{1}{4}\langle R_3\rangle^2_{\rho_{sep}}\right)^{\frac{1}{2}}.
\end{equation}
The inequality (\ref{Zhaoin}) is just the entanglement criterion for $ \mathbb{C}^2\otimes \mathbb{C}^d$ systems presented in \cite{Zhao2011}. It covers the corresponding result in \cite{Yu2003} for $d=2$. Hence, the separable criterion (\ref{Zhaoin}) is indeed a nonlinear improvement of $W_z^{U,V}$ by (\ref{thm}).

\section{Conclusions}

We have shown that any EW for $2\otimes d$ quantum systems can be improved by a nonlinear version.
The improved criterion has been shown to be strictly stronger than the original EW.
Any general, optimal or non-decomposable EW can always be nonlinearly improved, as shown by our examples. The method for entanglement detection given in \cite{Yu2003,Zhao2011} can be regarded as special cases of our approach. It is worthwhile to emphasize that, similar to an EW, its nonlinear improved criterion can be also easily verified by experimental measurements.
Compared with any original EW, the experimental implementation of the improved criterion only requires two more measurements on local observables. Besides,
it would be also interesting to investigate the nonlinear improvement of any EW for $m\otimes d$ ($m\ge 3$) quantum systems.

\section*{\bf Acknowledgments}
We thank the referee for the valuable comments and suggestions. This work is supported by NSFC (11775306, 11675113), the Fundamental Research Funds for the Central Universities (18CX02035A, 18CX02023A, 19CX02050A),
Beijing Municipal Commission of Education under grant No. KZ201810028042, and Beijing Natural Science Foundation (Z190005).


{\small \begin{thebibliography}{99}
\bibitem{applications} M.A. Nielsen and I.L. Chuang, \emph{Quantum computation and quantum information}, Cambridge University Press, Cambridge, UK, 2010; T. Gao, F.L. Yan, and Y.C. Li, Europhys. Lett. \textbf{84}, 50001 (2008); C.H. Bennett and S.J. Wiesner, Phys. Rev. Lett. \textbf{69}, 2881 (1992).

\bibitem{Gurvits2003-1} L. Gurvits, in \emph{Proceedings of the Thirty-Fifth Annual ACM
Symposium on Theory of Computing} (ACM Press, New York, 2003), pp. 10-19.

\bibitem{survey} R. Horodecki, P. Horodecki, M. Horodecki, and K. Horodecki, Rev. Mod. Phys. \textbf{81}, 865 (2009); O. G\"{u}hne and G. T\'{o}th, Phys.  Rep. \textbf{474}, 1 (2009); N. Friis, G. Vitagliano, M. Malik, and M. Huber, Nat. Rev. Phys. \textbf{1}, 72 (2019).

\bibitem{Terhal2001} B.M. Terhal, Linear Algebra Appl. \textbf{323}, 61 (2001).

\bibitem{Horodecki1996} M. Horodecki, P. Horodecki, and R. Horodecki, Phys. Lett. A \textbf{223}, 1 (1996).

\bibitem{Chruscinski2014} D. Chru\'{s}ci\'{n}ski and G. Sarbicki,
J. Phys. A: Math. Theor. \textbf{47}, 483001 (2014).

\bibitem{Gerke2018} S. Gerke, W. Vogel, and J. Sperling, Phys. Rev. X \textbf{8}, 031047 (2018); D. Chru\'{s}ci\'{n}ski, G. Sarbicki, and F. Wudarski, Phys. Rev. A \textbf{97}, 032318 (2018); M. Mozrzymas, A. Rutkowski,  and M. Studzi\'{n}ski, J. Phys. A: Math. Theor. \textbf{48}, 395302 (2015); A. Riccardi, D. Chru\'{s}ci\'{n}ski, and C. Macchiavello, arXiv:1901.08102v2 (2019).

\bibitem{Brandao2005} F.G.S.L. Brand\~{a}o, Phys. Rev. A \textbf{72}, 022310 (2005); J. Eisert, F.G.S.L. Brand\~{a}o, and K.M. Audenaert, New J. Phys. \textbf{9}, 46 (2007); O. G\"{u}hne, M. Reimpell, and R.F. Werner, Phys. Rev. A \textbf{77}, 052317 (2008).

\bibitem{MDIEW} C. Branciard, D. Rosset, Y.C. Liang, and N. Gisin, Phys. Rev. Lett. \textbf{110}, 060405 (2013); P. Xu, X. Yuan, L.K. Chen, H. Lu, X.C. Yao, X. Ma, Y.A. Chen, and J.W. Pan, Phys. Rev. Lett. \textbf{112}, 140506 (2014); J. Bowles, I. \v{S}upi\'{c}, D. Cavalcanti, and A. Ac\'{i}n, Phys. Rev. Lett. \textbf{121}, 180503 (2018); A. Mallick and S. Ghosh, Phys. Rev. A \textbf{96}, 052323 (2017); X. Yuan, Q. Mei, S. Zhou, and X. Ma, Phys. Rev. A \textbf{93}, 042317 (2016).

  \bibitem{EWphotons} M. Barbieri, F. De Martini, G. Di Nepi, P. Mataloni,
G.M. D'Ariano, and C. Macchiavello, Phys. Rev. Lett.
\textbf{91}, 227901 (2003); G. Lima, E.S. G\'{o}mez, A. Vargas, R.O. Vianna, and
C. Saavedra, Phys. Rev. A \textbf{82}, 012302 (2010); J. Dai, Y.L. Len, Y.S. Teo, B.G. Englert, L.A. Krivitsky, Phys. Rev. Lett. \textbf{113}, 170402 (2014); F. Brange, O. Malkoc, P. Samuelsson, Phys. Rev. Lett.  \textbf{118}, 036804 (2017).

\bibitem{Guhne2006-1} O. G\"{u}hne and N. L\"{u}tkenhaus, Phys. Rev. Lett. \textbf{96}, 170502 (2006).

\bibitem{Guhne2007-1} O. G\"{u}hne and N. L\"{u}tkenhaus, J. Phys.: Conf. Ser. \textbf{67}, 012004 (2007).

\bibitem{Moroder2008} T. Moroder, O. G\"{u}hne, and N. L\"{u}tkenhaus, Phys. Rev. A \textbf{78},
032326 (2008).

\bibitem{Arrazola2012} J.M. Arrazola, O. Gittsovich, and N. L\"{u}tkenhaus, Phys. Rev. A \textbf{85}, 062327 (2012).

\bibitem{Yu2005} S. Yu and N.L. Liu, Phys. Rev. Lett. \textbf{95}, 150504 (2005).

\bibitem{Guhne2006-2} O. G\"{u}hne, M. Mechler, G. T\'{o}th, and P. Adam, Phys. Rev. A \textbf{74}, 010301(R) (2006).

\bibitem{Zhang2007} C.J. Zhang, Y.S. Zhang, S. Zhang, and G.C. Guo, Phys. Rev. A \textbf{76}, 012334 (2007).

 \bibitem{2dsystem} B. Kraus, J.I. Cirac, S. Karnas, and M. Lewenstein, Phys. Rev. A \textbf{61}, 062302 (2000); A. Singh, A. Gautam, and K. Dorai, Phys. Lett. A \textbf{383}, 1549 (2019).

\bibitem{Yu2003} S. Yu, J.W. Pan, Z.B. Chen, and Y.D. Zhang, Phys. Rev. Lett. \textbf{91}, 217903 (2003).
\bibitem{Zhao2011} M.J. Zhao, T. Ma, S.M. Fei, and Z.X. Wang, Phys. Rev. A \textbf{83}, 052120 (2011).

\bibitem{Goldberg1982} M. Goldberg and E. Tadmor, Linear Algebra Appl. \textbf{42}, 263 (1982).
\bibitem{Horn1985} R.A. Horn and C.R. Johnson, \emph{Matrix analysis}, Cambridge University Press, Cambridge, UK, 1985.
\bibitem{Lewenstein2000} M. Lewenstein, B. Kraus, J.I. Cirac, and P. Horodecki, Phys. Rev. A \textbf{62}, 052310 (2000).
\bibitem{Horodecki1997} P. Horodecki, Phys. Lett. A \textbf{232}, 333 (1997).
\end{thebibliography}}
\end{document}